\documentclass[journal]{IEEEtran}

\IEEEoverridecommandlockouts

\usepackage{cite}
\usepackage{url}
\usepackage{graphicx}
\usepackage{amsmath}

\usepackage{amssymb}
\usepackage{color}
\usepackage{bm}
\usepackage{multirow}
\usepackage{algorithm}
\usepackage{algpseudocode}
\usepackage{todonotes}

\usepackage{rotating}

\usepackage{varwidth}

\usepackage{algorithm}
\usepackage{algpseudocode}
\algnewcommand\algorithmicinput{\textbf{Input:}}
\algnewcommand\INPUT{\item[\algorithmicinput]}
\algnewcommand\algorithmicoutput{\textbf{Output:}}
\algnewcommand\OUTPUT{\item[\algorithmicoutput]}

\usepackage{mathtools}

\DeclareMathAlphabet\mathbfcal{OMS}{cmsy}{b}{n}

\graphicspath{{figures/}}
\usepackage{subcaption}

\renewcommand{\epsilon}{\varepsiolon}
\renewcommand{\phi}{\varphi}

\def\BibTeX{{\rm B\kern-.05em{\sc i\kern-.025em b}\kern-.08em
    T\kern-.1667em\lower.7ex\hbox{E}\kern-.125emX}}

\begin{document}


\title{ Virtual Sectorization to Enable \\ Hybrid Beamforming in mm-Wave mMIMO  
\thanks{The research was carried out at Skoltech and supported by the Russian Science Foundation (project no. 24-29-00189).
}}

\author{Roman~Bychkov,
        Andrey~Dergachev,
        Alexander~Osinsky,
        Vladimir~Lyashev,
        and Andrey~Ivanov
        \thanks{R. Bychkov, A. Dergachev, A. Osinsky, and A. Ivanov are with Skolkovo Institute of Science and Technology, Russia. E-mails: 
        R.Bychkov@skoltech.ru,
        Andrey.Dergachev@skoltech.ru, alexander.osinsky@skoltech.ru, 
        an.ivanov@skoltech.ru}
        \thanks{V. Lyashev is with Moscow Institute of Physics and Technology, Moscow, Russia. E-mail: lyashev.va@mipt.ru.}
        }

\maketitle
\IEEEdisplaynontitleabstractindextext

\begin{abstract}
Hybrid beamforming (HBF) is a key technology to enable mm-wave Massive multiple-input multiple-output (mMIMO) receivers for future-generation wireless communications. It combines beamforming in both analog (via phase shifters) and digital domains, resulting in low power consumption and high spectral efficiency. In practice, the problem of joint beamforming in multi-user scenarios is still open because an analog beam can't cover all users simultaneously. 

In this paper, we propose a hierarchical approach to divide users into clusters. Each cluster consists of users inside a virtual sector produced by the analog beamforming of an HBF-based mMIMO receiver. Thus, inside each sector, a lower-cost digital beamforming serves a limited number of users within the same cluster.

Simulations with realistic non-line-of-sight scenarios generated by the QuaDRiGa 2.0
demonstrate that our methods outperform standard FFT-based alternatives and almost achieve SVD-based beamspace performance bound. 

 \vskip 0.2cm


\end{abstract}


\section{Introduction}
\label{sec:introduction}

\IEEEPARstart {E}{nhanced} Mobile Broadband (eMBB) \cite{5Gview}, which provides a fast mobile Internet, is the primary research area of cellular networks. Using a wide frequency band, improving the signal-to-noise ratio, and implementing the technology of so-called spectrum ``reuse'' are the main tools to increase the data rate. The spectrum ``reuse'' is the key idea of the Massive Multiple Input Multiple Output (M-MIMO) technology that presumes separating data streams in the spatial domain \cite{Ahmed}.

Because the spectrum width is constrained by several quite narrow frequency bands and the number of antennas is constrained by the design of the radio unit, it is already clear that these approaches are at their limit inside the 5G communications. As a result, moving to the upper millimeter waves, where a substantially wider bandwidth is available for data transmission, makes sense for systems of the next generation, such as 6G cellular communication systems. A conventional form factor can accommodate many times more antennas if the carrier frequency is increased up to tens of gigahertz and the antenna size is decreased. Thus, the efficiency of the spectrum reuse will increase, and it makes sense to employ an extremely large number of antennas (Ultra Massive MIMO, or UM-MIMO) in the upper millimeter range. 

Major challenges in the implementation of numerous antenna elements are their high cost and energy consumption. Compared to the previous generations, the number of antennas is increased from $8$ in 4G, to $1024$ in 5G, resulting in at least 
\begin{equation}\label{complexity}
\textbf{growth in complexity = }[\frac{1024}{8}]^2=128^2=16384
\end{equation}
times increase in the complexity of channel estimation and signal detection algorithms \cite{osinsky2020theoretical, HighPerform}. Moreover, the Common Public Radio Interface (CPRI) \cite{Kalfas2019} has insufficient capacity to transfer so many digital antenna signals from the remote radio head (RRH) to the baseband unit (BBU). 

Thus, the receiver's huge computational complexity and power consumption are the key obstacles to installing an ultra-large number of antennas. Because of this, scaling up already existing signal processing techniques is practically impossible when implementing a new system.

\subsection{ Hybrid beamforming}

There are high expectations from the use of millimeter (mm) waves in 5G because this bandwidth is vacant \cite{5Gview}, but path loss for mm waves is dramatic and should be compensated by the use of much larger antenna arrays \cite{nguyen2023deep}. A fully digital architecture is not feasible in this case since the digital-to-analog converters consume too much energy, and the selection of beamspace in a hybrid architecture is an open problem. 

A promising solution to the high complexity problem lies in the hybrid beamforming (HBF) technology \cite{Molisch, Ahmed, Kalinin}, which employs analog beamforming in the remote radio head (RRH) together with digital beamforming in the baseband unit (BBU). The HBF technology allows reducing the number of digital ports by using analog phase shifters. It is aimed to guarantee the channel gain at acceptable complexity in scenarios where the distribution of target users is sparse in the spatial domain \cite{phy_hbf}. Thus, the HBF in mMIMO systems is intended to maximize the sum rate that can reach the fully digital beamforming system performance at a cost slightly higher than the fully analog beamforming system. Therefore, a combination of mMIMO systems, mmWave, and HBF techniques can achieve higher data rates and cell coverage in future generations of wireless networks. To sum up, it requires less hardware complexity, energy consumption, and cost by minimizing the number of analog-to-digital (ADC) and digital-to-analog (DAC) converters at the base station. 

There are several papers focusing on the mmWave Massive MIMO based on HBF. A deep survey is given in \cite{Ahmed}, where the authors focus on the implementation, signal processing, and application aspects of the HBF. For each feature, the authors gave an overview of the current state-of-the-art research and discussed the challenges (multi-user HBF) and problems (relation to the standardization activities). Here we are mostly interested in how one can train phases for phase shifters and select the appropriate digital beamspace for the hybrid architecture. In \cite{precoding} the authors suggested adapting the phases iteratively with an algorithm similar to successive interference cancellation. In \cite{kronecker} the task was split as a Kronecker product of phase shifts. Then some of the parameters (with a dimension equal to the interference dimensionality) were used to compensate for interference and others were used to align with the optimal estimated signal directions. Finally, one can also construct a more general problem, as if phase shifters could also update the amplitudes, but only use the phase part of the solution \cite{ampl1,ampl2}.

Unfortunately, the calculation of the steering vector (precoding \cite{lyashev} at the transmitter or weighing at the receiver) is an open problem for hybrid beamforming. The problem occurs in multi-user scenarios when one analog beam can't cover users distributed in the spatial domain without degradation in radiation pattern gain.

In this paper, we examine user clustering methods in application to the sectorization problem in HBF. Because HBF  provides a trade-off between performance and computational complexity, it stands as the most promising technology for the design of cellular communication systems with an extremely high number of antennas.

\subsection{Our contribution}

The major problem of the HBF technology is the beamforming algorithm design for multi-user scenarios. In such cases, the analog beam should cover all users in the spatial domain simultaneously to avoid performance degradation. Obviously, in most multi-user scenarios, we can't process all users simultaneously using a single high-gain analog beam without losses. In this case, we should justify the minimal reuse factor, in other words, to answer the question: how many analog beams do we need in a multi-user scenario ? Or how to calculate these beams for a given reuse factor.  

To solve the aforementioned problem, we propose a new clustering algorithm for the partially-connected (PC) HBF architecture under the use of OFDM and uplink operation. The algorithm presumes user clasterization for the analog beams and a fixed fast Fourier transform (FFT) for digital beamspace selection \cite{lower_bound}.

Proposed algorithms were tested in a realistic propagation channel generated with QuaDRiGa 2.0 \cite{Quadriga} simulation tool. To the best of our knowledge, it is the first time unsupervised clustering has been tested on partially connected structures in a practical environment.

\section{System model and algorithms}

Strict conditions for a future HBF architecture are placed on the complexity growth derived in eq. \eqref{complexity}. There are several HBF architectures considered in the literature: fully connected and partially connected, statically and dynamically connected \cite{Ahmed}. 

\begin{figure}[t!]
\centering
\includegraphics[width=0.95\columnwidth]{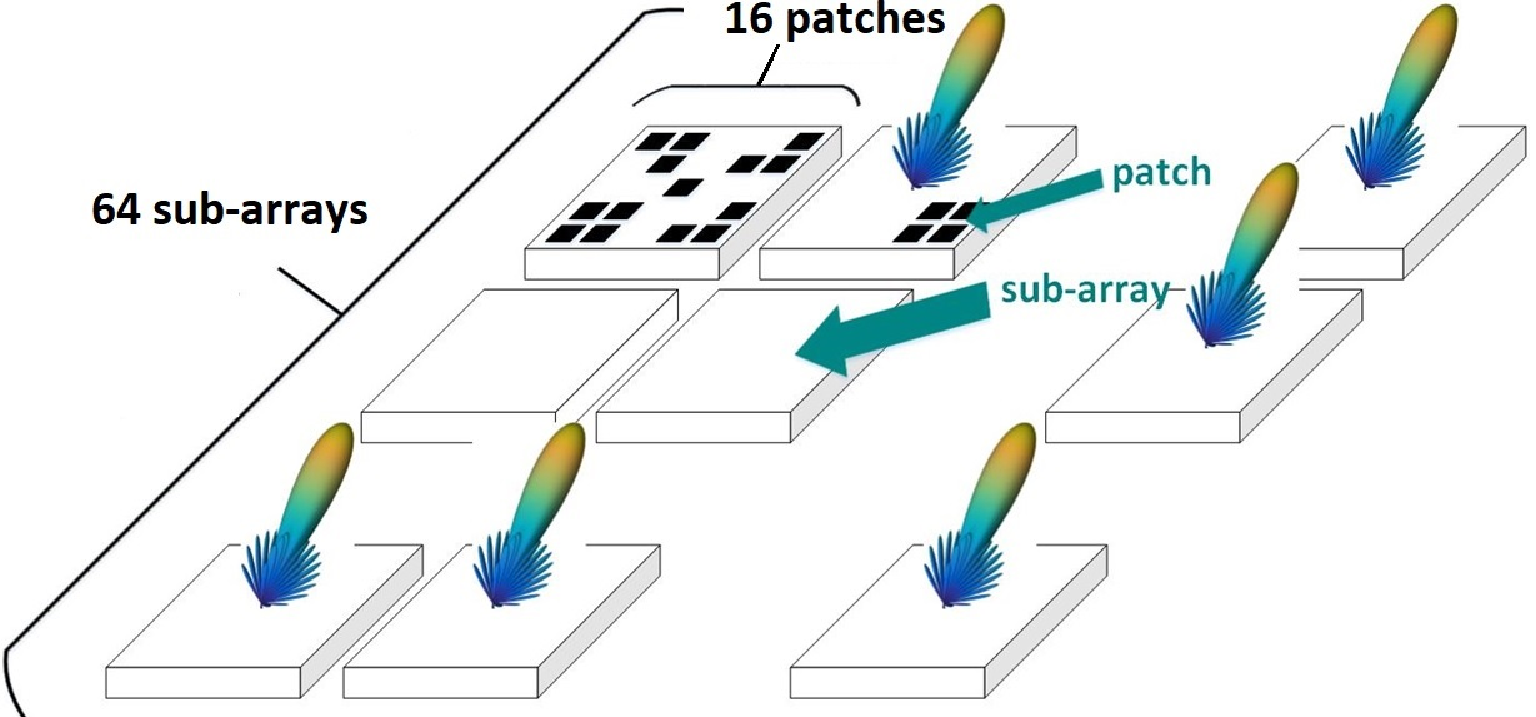}
\caption{Sub-array HBF.}
\label{fig:HBF}
\end{figure}

The HBF complexity must be incredibly low, therefore, we focus mainly on static partially connected architecture as having the lowest complexity among other architectures, and thus, be the most attractive in practice. The partially connected (PC), also known as subarray-based or sub-connected, hybrid beamforming architecture \cite{allphases,twophases:absolutevalue} has been chosen as the most promising thanks to its low complexity. It is shown in Fig. \ref{fig:HBF}. The corresponding beamforming scheme is presented in Fig. \ref{fig:HBF2}.  

\subsection{Analog beamforming (phase shifters)}

In SBHB architecture, each digital channel is connected to a subarray of antenna elements via dedicated phase shifters. Each receiving antenna is equipped with a phase shifter, and all $N_{RX}$ antennas are divided into $M$ subarrays. Inside each subarray, the sum of phase-shifted analog signals is calculated and further sampled, thus implementing an analog beamforming. There also exist more complicated hybrid architectures, when each antenna can be connected to multiple phase shifters \cite{allphases,twophases:absolutevalue}.

\subsection{Beamspace selection}

\begin{figure}[t!]
\centering
\includegraphics[width=0.95\columnwidth]{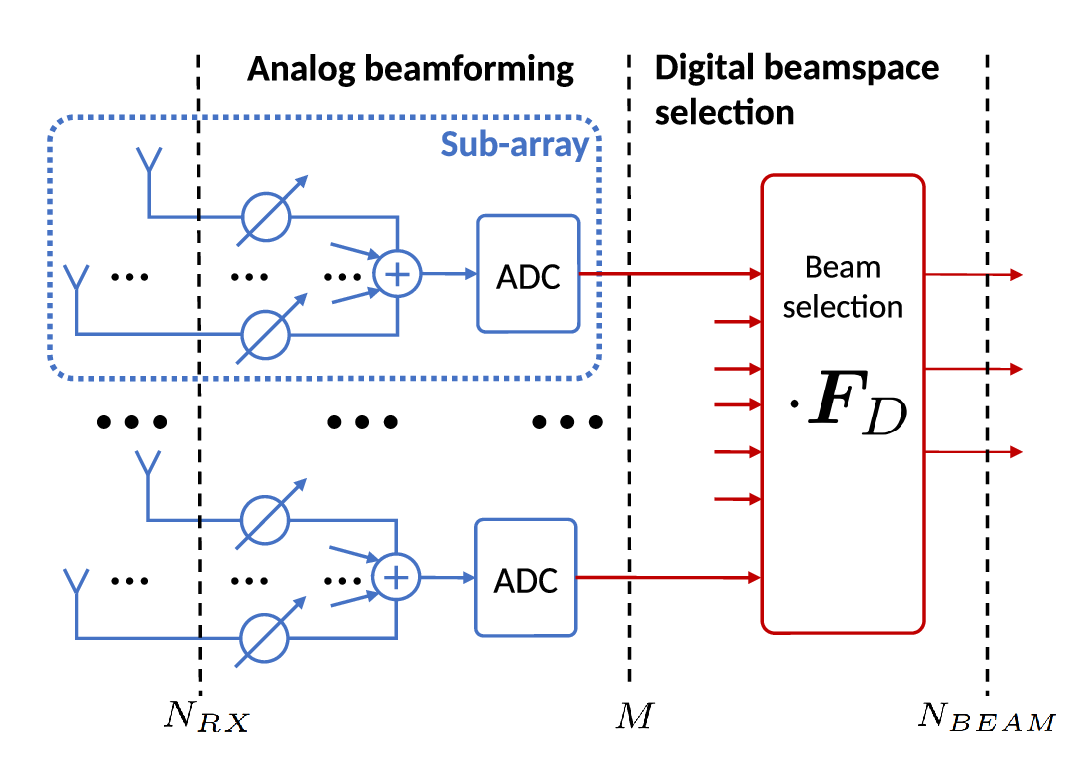}
\caption{Double beamforming HBF scheme at the receiver.}
\label{fig:HBF2}
\end{figure}

A further way of the complexity reduction consists of $2$ subsequent steps implemented in the digital domain: transformation to the beamspace with a further selection of a low dimension subspace, as shown in Fig. \ref{fig:HBF2}. After transforming the antenna signal to the beamspace, it becomes sparse: compared with omnidirectional antennas, beams shape an orthogonal space. Thus, by choosing the subspace containing the user signals, we exclude noisy beams. Therefore, signal dimensionality is reduced again, resulting in less computational complexity and providing better accuracy due to filtering out noisy subspaces \cite{access}. The best possible low-dimensional approximation is provided by singular value decomposition (SVD). However, the complexity of decomposition itself and the corresponding transformation is prohibitively large for 5G. A common solution is to use an FFT, in which columns represent a set of uniformly distributed directions. Unfortunately, the FFT-based beamspace selection accuracy is worse than the SVD-based one because the actual propagation signal taps are not fully aligned with the fixed FFT beams.  

\subsection{Digital beamforming}

\begin{figure}[t!]
 \centering
 \includegraphics[width=0.90\columnwidth]{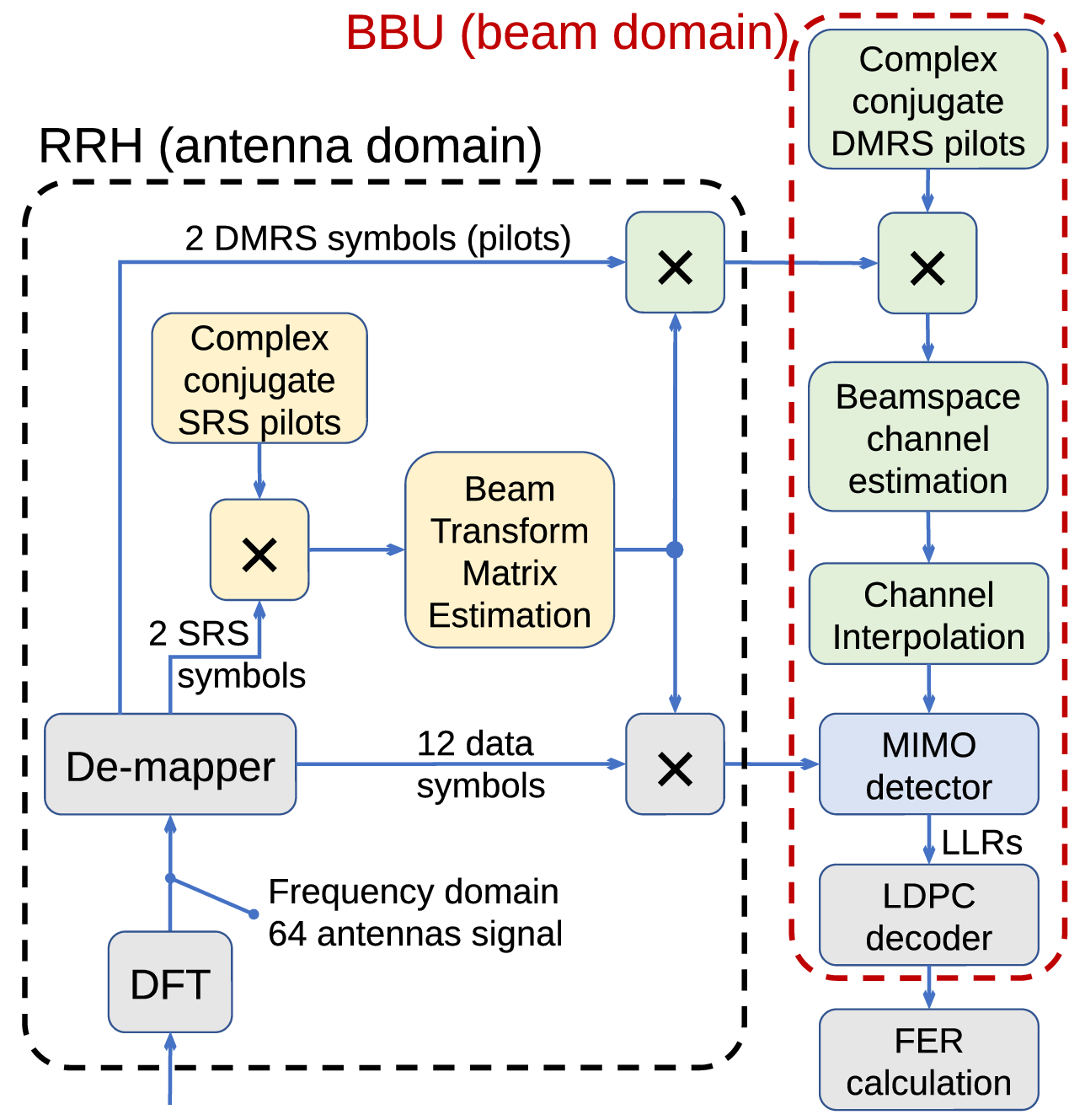}
 \caption{
 Receiver structure with the beamspace transformation.
 }
 \label{beamspace}
 \end{figure}

A simplified digital workflow of the receiver is shown in Fig. \ref{beamspace}. Firstly, the digital signal from each subarray is transformed to the frequency domain. Then, the De-mapper extracts the Sounding Reference Signal (SRS), Demodulation Reference Signal (DMRS) pilots, and data symbols. The SRS is employed to estimate the optimal beamspace transformation matrix at the RRH. This matrix is then applied to convert the digital subarray signal to the beamspace domain with the dimensionality $N_{BEAM}$ less than $M$. Then the data is transferred to the BBU, where the linear MIMO detection \cite{cloud}, \cite{HighPerform} performers the final digital beamforming in the beamspace domain.

\subsection{Signal model}

The linear MIMO detection is assumed as the baseline for signal detection \cite{cloud} thanks to its lowest complexity compared to other detector types. It is universal in applications, demonstrates the best performance in a majority of application scenarios, and has no tunable parameters. 

In contrast, joint analog beamforming and beamspace selection algorithms lack better performance in practice. To examine them in realistic scenarios, define the received signal in the beamspace domain (before MIMO detection) $\bm y_{HBF}$ as follows:
\begin{equation}
\begin{aligned}
    \bm y_{HBF} &= \bm F_{D} \bm F_{A}[ \bm H \bm x + \bm w] \\
    &=  \tilde{ \bm H } \bm x +  \tilde {\bm w},
\end{aligned}
\end{equation}
where $\bm H \in \mathbb{C}^{N_{RX}\times N_{TX}}$ is the propagation channel matrix, 
$\bm x  \in \mathbb{C}^{N_{TX}}$ is the user signals vector,
$\bm w  \in \mathbb{C}^{N_{TX}}$ is the additive white Gaussian noise (AWGN),
$\bm F_{A} \in \mathbb{C}^{M \times N_{RX}}$ is the radio frequency (analog) beamforming matrix,
$\bm F_{D} \in \mathbb{C}^{N_{BEAM} \times M}$ is the baseband (digital) beamspace selection matrix,
$ \bm {\tilde H}$ and $\bm {\tilde w}$ are the corresponding effective (beamspace) channel and the noise matrices.

 In PC HBF architecture, the analog beamforming matrix $\bm F_{A}$ is defined as


\begin{equation}\label{eq:FA}
\begin{aligned}
    \bm F_{A} &= 
\left[ \begin{matrix}
  e^{i\phi_{11}} & \dots & e^{i\phi_{1L}}& 0           & \dots & 0                  & \dots\\
  0              & \dots & 0          & e^{i\phi_{21}} & \dots & e^{i\phi_{2L}}     & \dots\\
  \vdots         & \vdots& \vdots     & \vdots         & \ddots & 0                  & \dots\\
  0              & \dots & 0          & 0              & \dots & 0                   & \dots\\
\end{matrix}\right.\\
\\
&\qquad
\left. \begin{matrix}
  \dots                       & 0           & \dots & 0 \\
  \dots                       & 0           & \dots & 0\\
  \dots                       & \vdots                      & \ddots & 0\\
  \dots                       & e^{i\phi_{M1}}              & \dots & e^{i\phi_{ML}}\\
\end{matrix} \right]
\end{aligned}
\end{equation}

where $\phi_{ij}$ is the phase state at $j$-th phase shifter of the $i$-th subarray, $L=N_{RX}/M$ is the number of phase shifters employed at each individual subarray.

\section{Proposed algorithm}

For a single-user case, the phase adjustment is pretty simple. 

Assume a line-of-sight channel as known, and $\bm u_i$ is the first singular vector (or line-of-sight direction steering vector) of the single user channel matrix $\bm H_i$, here $i$ is the user index. Therefore, to maximize the received signal power, all subarrays should implement the beamforming with vector $\bm u_i$. Unfortunately, phase shifters can adjust only the corresponding phase as:

\begin{equation} \label{subarray_direction}
    \bm {\phi}_{m} = \angle \bm u_i^m, 
\end{equation}
where $\bm {\phi}_{m}=[\phi_{m1}\quad\phi_{m2}\quad...\quad\phi_{mL}]$ is the vector of phase states for the $m$-th subarray. Thus, for the $m$-th subarray, instead of the SVD-BF $\bm u_i^m$ we employ its approximation $e^ {\sqrt{-1}\angle \bm u_i^m}$ with a small loss in performance.



\begin{figure}[t!]
\centering
\includegraphics[width=1\columnwidth]{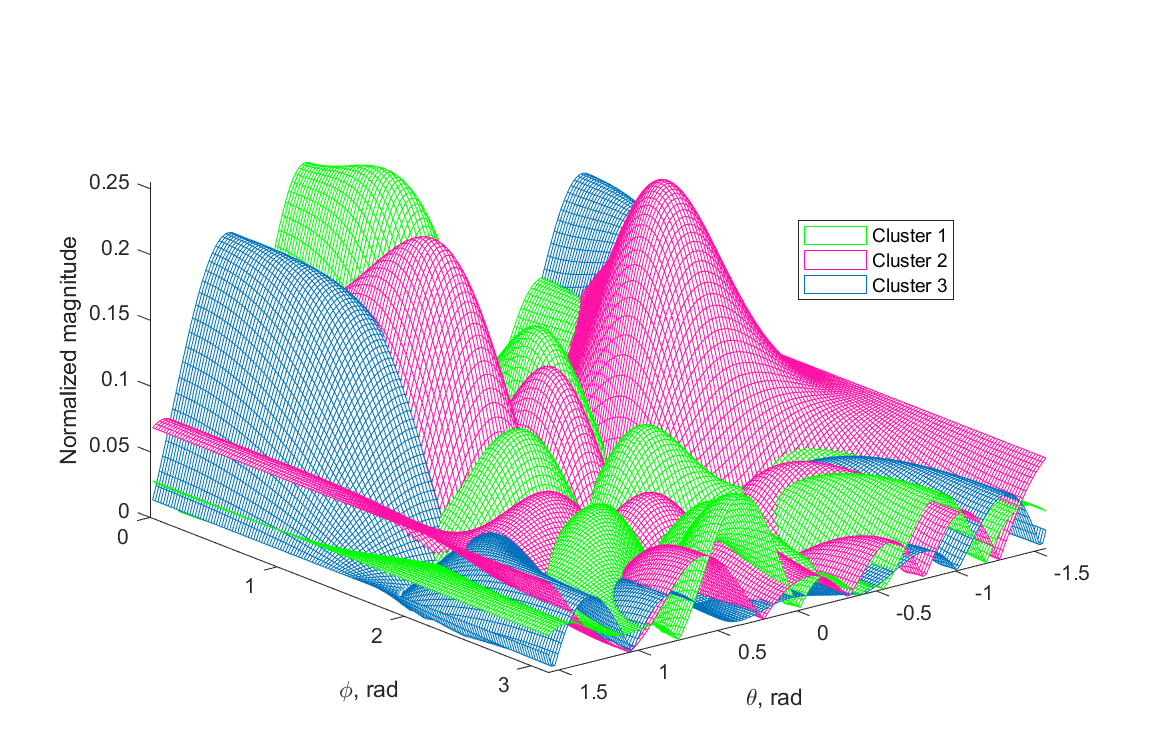}
\caption{
Subarrays radiation pattern for 3 clusters.
}
\label{fig:pattern}
\end{figure}

The case of multi-user beamforming is more complicated. PC HBF imposes strict restrictions on angles-of-arrival (AoA) for users inside an analog beam. Generally, to provide a high beamforming gain, all subarrays should have approximately the same phase state $\bm {\phi}_{m}$ corresponding to the AoA of user signals, but in practice it is not exactly true and, as we will be showing, such assumption leads to performance loss.

\textit{Example:} suppose users have a significant variation in their angles of arrival (AoA). Therefore, to process them there are $2$ approaches: set a similar phase states on all subarrays corresponding to omnidirectional antenna or set phase states on each subarray depending on only one user AoA. In approach 1 the analog beamforming gain is absent, therefore, this beamforming is the worst. In approach 2 the analog beamforming gain is guaranteed for each user, but the user signal will be presented in a limited number of subarray outputs, i.e. the user power will be lost in the digital domain.

\subsection{Motivation for user clustering}

In practice, user distribution can be any, but, definitely, it is non-uniform and real environment-dependent. For example, users have concentrated alone on roads, nearby shopping centers, buildings, and so on. Probably, the overall distribution is sparse in the spatial domain \cite{bychkov2021data}. Thus, one can join spatially correlated users into $K$ clusters so that all users inside each $k$-th cluster have negligible attenuation inside a receiving pattern defined by an existing vector $e^{\sqrt{-1}\angle \bm u_k}$, i.e. a cluster center. In other words, we try to fit users into a limited number of analog beams having a high gain. 

Having a limited number of clusters allows for predicting the required number of analog beams (hybrid multi-beamforming) or the analog beam reuse factor. 

Therefore, the goal is to find user clustering resulting in the minimal performance loss for a given number of clusters $K$. If this loss is acceptable for the system, the PC HBF is valid, and the complexity gain compared to the fully digital BF can be calculated.

\subsection{Hierarchical clustering } 

To achieve user partition, we propose using the unsupervised hierarchical clustering procedure. The hierarchical clustering for HBF was originally proposed in \cite{Xu2018}. 

In the beginning, each user $i$ is treated as a separate cluster $g_i$ of size 1, $\left| g_i \right| = 1$. Then, at each step, two clusters with the least distance between them are collapsed into a single cluster. 

The distance between clusters $g_k$ and $g_l$ is determined as the maximum distance between users in these clusters:
\begin{equation}\label{eq:cluster_dist}
    d_{C} (g_k, g_l) = \max\limits_{i \in g_k, k \in g_l} d_{U} (i,j),
\end{equation}

where

\begin{equation} \label{array_dist}
    d_{U}(i,j) = 1 - \left| \bm u_i^H \cdot \bm u_j \right|.
\end{equation}

 is the distance between $2$ users. 
 
 The algorithm proceeds until the predefined number of clusters $K$ is achieved.

\subsection{Hierarchical clustering-PC }

Thus, the distance function should be adjusted taking into account the subarray structure. Let's use the following function to estimate the similarity of two users $i$ and $j$:

\begin{equation}\label{subarray_dist}
    d_{U}(i,j) = 1 - \sum_{m=1}^{M} \left| [\bm u^m_i]^H \cdot \bm u^m_j \right|,
\end{equation}
where $i,j$ are user indexes, $m$ is a subarray index, $\bm u^m_i$ is the channel vector of $i$-th user at $m$-th subarray.

\subsection{Cluster center}

For each cluster, we have to choose the corresponding ``center''. This center represents the AoA vector, which is the same for each user inside the cluster. The received power is the major criterion, therefore 
let us define
\begin{equation} \label{subarray_dir}
  \bm u_{k} = \operatorname{svd}_1 \left( \bm H_k \right), \quad \bm u_{k} \in \mathbb{C}^{N_{RX}},
\end{equation}
as the first left singular vector in the singular value decomposition of $\bm H_k$ (corresponding to the largest singular value).
Here $\bm H_k \in \mathbb{C}^{N_{RX} \times \left| g_k \right|}$ is the channel matrix consisting of only users inside $k$-th cluster (number of columns is equal to the number of users in the cluster). Vector $\bm u_{k}$ is then used to select the phase shifts.






The end-to-end workflow for PC HBF user clustering is presented at Algorithm \ref{alg:clustering}.

\begin{algorithm}[t!]
\caption{\textbf{Hierarchical clustering-PC}}\label{alg:clustering}
\begin{algorithmic}[1]

\INPUT User directions $\bm U = \{ \bm u_1, \bm u_2,\dots, \bm u_{N_{TX}}\}$;
\OUTPUT User clusters $G=\{g_1, g_2,\dots, g_K\}$, \hfill \break phases $\{\phi_{m,l}\}_k$  (see \eqref{eq:FA}). \\
Combine users into clusters using the \textbf{hierarchical method-PC} according to eqs. \eqref{subarray_dist} and \eqref{eq:cluster_dist}. The distance between two clusters is estimated as the maximal distance between pairs of users inside the clusters. \\
Determine the cluster directions using \eqref{subarray_dir}. Calculate phase states using \eqref{subarray_direction}. Set phase shifters accordingly.

\end{algorithmic}
\end{algorithm}

\subsection{Hierarchical clustering-PC, multi-center} 
To justify the choice for the cluster center defined by eq. \eqref{subarray_dir}, we examine the multi-center version of \textbf{Hierarchical clustering-PC}. In \textbf{Hierarchical clustering-PC-MC} instead of setting all subarrays to a cluster center, each subarray adjusts the phase shifters according to the $1$-st SVD vector \eqref{subarray_direction} of a user inside this cluster. For example, assume the cluster consists of $2$ users. Thus, the first $M/2$ subarrays will be focused on the $1$-st user, and the second $M/2$ subarrays will be focused on the $2$-nd user. Therefore, subarray beams inside the cluster are both subarray-dependant and user-dependent, but each user is detected using all $M$ subarrays.

\section{Simulation results}
\label{sec:experiments}

For simulations, we employed the Quadriga 2.0 software \cite{Quadriga} to generate a realistic LOS radio channel focusing on $6G$. Assume that the propagation channel is ideally estimated. The linear MMSE algorithm \cite{cloud} was applied for signal detection. Other system parameters are defined in Table \ref{tab:params}. 


\subsection{Other existing algorithms}

\textbf{Max power-1} adjusts all subarrays for all users at once in a way that the total power is maximized. Its center can be found with an SVD of the full channel matrix $\bm H \in \mathbb{C}^{N_{RX} \times N_{TX}}$ as 
\begin{equation} \label{apriori_dir}
    \bm u = \operatorname{svd}_1 \left( \bm H \right), \quad \bm u \in \mathbb{C}^{N_{RX}},
\end{equation}
and subarrays are set to $e^{\sqrt{-1}\angle \bm u}$ states. This approach doesn't divide users into clusters. Its performance demonstrates the necessity of user clustering because the subarray beam tends toward an omnidirectional antenna.

\textbf{Max power-K} also employs the SVD of the full channel matrix $\bm H \in \mathbb{C}^{N_{RX} \times N_{TX}}$ fo find the cluster centers: 
\begin{equation} \label{apriori_dir}
    \bm U = \operatorname{svd}_K \left( \bm H \right), \quad \bm U \in \mathbb{C}^{N_{RX} \times K},
\end{equation}
but now the $K$ largest left singular vectors are selected. This algorithm can be interpreted as the clustering based on $K$ most powerful directions. Then we assign each user to a cluster having the minimal distance between user and center \eqref{subarray_dist}.


\textbf{Fully digital} presumes the antenna signal sampling without analog beamforming, i.e. there are no phase shifters, and each antenna signal is sampled by ADC. Thus, $N_{RX}=M$, and no clusters are required. The fully digital architecture is considered as the upper bound for the system performance.




\begin{table}[t!]
    \renewcommand{\arraystretch}{1.4}    
        \begin{tabular}{|p{2.3cm}|p{1.0cm}||p{2.0cm}|p{1.8cm}|}\hline
            \textbf{Parameter}    & \textbf{Value}  &  \textbf{Parameter}& \textbf{Value}   \\ \hline
            \textbf{Carrier frequency}   &  50GHz  &  \textbf{Modulation} & QAM16 \\ 
            \textbf{Subcarrier spacing}  &  120kHz   &  \textbf{N of scenarios} & 140 \\ 
            \textbf{BS height}  & 25m               &  \textbf{N of noise seeds} & 16 \\ 
            \textbf{UE height} &  1.5m              &  \textbf{RB number} & 4 \\ 
            \textbf{Vertical antennas spacing} & 0.9$\lambda$   &  \textbf{Decoder} & LDPC (144,288)  \\ 
            \textbf{Horizontal antennas spacing} & 0.5$\lambda$ &   \textbf{Channel model}  & 3GPP-3D, Berlin, Dresden \\
            \textbf{N of BS antennas}   &   1024  & \textbf{Channel type}   &  LOS  \\ 
            \textbf{N of UE antennas}  &   2 & \textbf{User speed} & 5 km/h \\ 
            \textbf{UE number}  &   240 & \textbf{$\bm{N_{beam}}$} & 32 \\ 
            \textbf{N-of-clusters}  &   8 &  &  \\ 
            \hline
        \end{tabular}
        \caption{Simulation parameters.}
        \label{tab:params}
\end{table}

\subsection{Discussion}

\begin{figure}[t!]
\centering
\includegraphics[width=0.99\columnwidth]{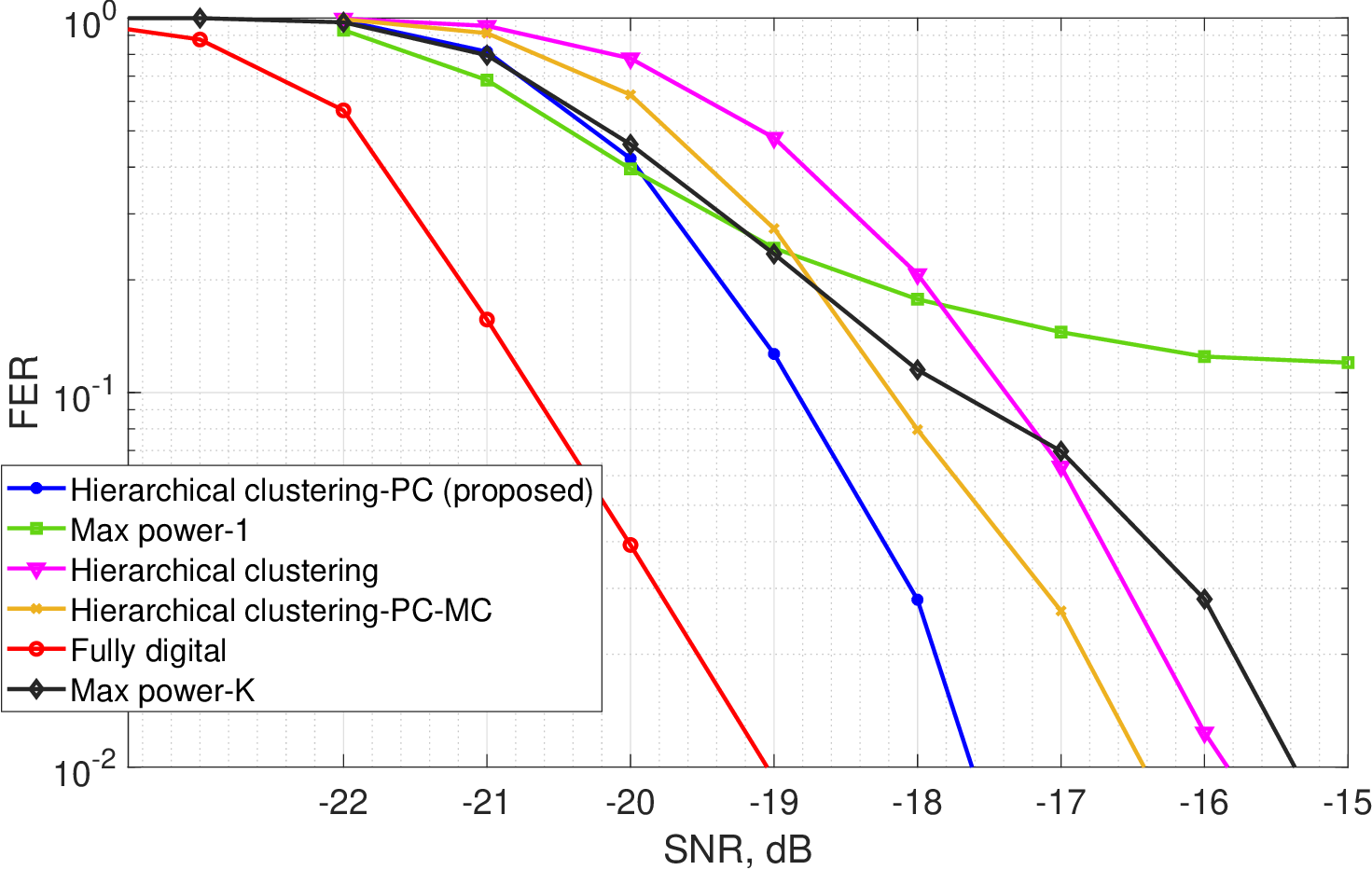}
\caption{Frame error rate (the ratio of codewords received with errors to total received codewords).
}
\label{FER}
\end{figure}

The frame error rate (FER) results are presented in figure \ref{FER}. The proposed \textbf{Hierarchical clustering-PC} outperforms the \textbf{Hierarchical clustering-PC-MC} by $1.2dB$. In \textbf{Hierarchical clustering-PC-MC} approach, the user signal loss can be significant, since focusing subarrays on a particular user can lead to other users falling onto the edge of the subarray radiation pattern.
Therefore, the beamforming using the cluster center is preferable than the using per user beamforming for separate subarrays inside this cluster.

Then, proposed \textbf{Hierarchical clustering-PC} also outperformed the baseline version \textbf{Hierarchical clustering} algorithm by almost $1.8dB$. In \textbf{Hierarchical clustering} approach, only highly-correlated users are grouped together, thus making almost impossible their separation in the digital domain. Therefore, the user correlation should be calculated per subarray in the eq. \eqref{subarray_dist} of distance calculation. Therefore, the AoA strongly depends on the subarray index. Thus, the necessity of distance calculation with subarray structure is confirmed. 

Compared to \textbf{Fully digital} antenna array, the loss is almost $1.4dB$. Therefore, the algorithm \textbf{Hierarchical clustering-PC} is a reasonable replacement of the \textbf{Fully digital} solution. 

The reason of \textbf{Max power-1} loss is clear: extremely low gain of subarray beamforming and inability to process all users in only one cluster because of a limited width of the radiation pattern. 

\textbf{Max power-K} approach, firstly, has the same disadvantage as \textbf{Hierarchical clustering}. Moreover, each cluster center is calculated over all users, but futher it is employed to only a few users that belong to this cluster. This strongly degrades the accuracy of the cluster center calulation, and, as a result, detection performance.        

\section{Conclusion}
\label{sec:conclusion}

Proposed algorithm \textbf{Hierarchical clustering-PC} is a reasonable alternative of the \textbf{Fully digital} beamforming. Using $K=8$ clusters, the number of required ADC is reduced from $N_{RX}=1024$ down to $M=64$. Simulations with realistic $6G$ LOS channel generated by the Quadriga 2.0 demonstrate only $1.4$dB performance loss. 

In practice, sectorization can be implemented offline to find K analog beams (phase shifters state + center for each cluster) for each base station. Each cluster means a fixed sector of the station. Then, in the online mode, each user is assigned to a sector (cluster) according to the minimal distance \eqref{subarray_dist} to the cluster center. As a result, the base station can consist of $K$ fixed sectors, in each of which a common size-$M$ flexible digital beamforming is utilized to detect the in-sector users.

\section{Acknowledgment}

The authors acknowledge the use of computational cluster Zhores \cite{Zhores2019} for obtaining the results presented in this letter.

\bibliographystyle{IEEEtran}
\bibliography{main.bib}

\end{document}